\documentclass[10pt,letterpaper]{article}
\usepackage[top=0.85in,left=2.75in,footskip=0.75in]{geometry}

\usepackage{amsmath,amssymb}
\usepackage[textsize=tiny]{todonotes}

\usepackage{xfrac}
\usepackage{float}

\usepackage{changepage}

\usepackage{textcomp,marvosym}

\usepackage{cite}
\usepackage{url}
\usepackage{nameref,hyperref}

\usepackage[right]{lineno}
\usepackage{upgreek}

\usepackage[T1]{fontenc}
\usepackage{lmodern}

\usepackage[nopatch=eqnum]{microtype}
\DisableLigatures[f]{encoding = *, family = * }

\usepackage{xcolor}

\usepackage{array}

\newcolumntype{+}{!{\vrule width 2pt}}

\newlength\savedwidth

\raggedright
\usepackage[aboveskip=1pt,labelfont=bf,labelsep=period,justification=raggedright,singlelinecheck=off]{caption}

\makeatletter
\renewcommand{\@biblabel}[1]{\quad#1.}
\makeatother

\usepackage{lastpage,fancyhdr,graphicx}
\usepackage{epstopdf}
\fancyheadoffset[L]{2.25in}
\fancyfootoffset[L]{2.25in}
\begin{document}
\vspace*{0.2in}

\begin{flushleft}
{\Large
\textbf\newline{A method for site-specifically tethering the enzyme urease to DNA origami with sustained activity} 
}
\newline
\\
Ian Murphy\textsuperscript{1},
Keren Bobilev\textsuperscript{1},
Daichi Hayakawa\textsuperscript{1},
Eden Ikonen\textsuperscript{1},
Thomas E. Videb{\ae}k\textsuperscript{1}
Shibani Dalal\textsuperscript{1},
Wylie W. Ahmed\textsuperscript{2,3,4},
Jennifer L. Ross\textsuperscript{5},
W. Benjamin Rogers\textsuperscript{1}
\\
\bigskip
\textbf{1} Martin A. Fisher School of Physics, Brandeis University, Waltham, MA 02453 USA
\\
\textbf{2} Laboratoire de Physique Théorique (LPT), Université de Toulouse, CNRS, UPS, 31400 Toulouse, France
\\
\textbf{3} Molecular, Cellular and Developmental biology unit (MCD), Centre de Biologie Intégrative (CBI), Université de Toulouse, CNRS, UPS, 31400 Toulouse, France
\\

\textbf{4} Department of Physics, California State University, Fullerton, CA 92831 USA
\\
\textbf{5} Department of Physics, Syracuse University, Syracuse, NY 13244 USA
\\
\bigskip

\end{flushleft}
\section*{Abstract}
Attaching enzymes to nanostructures has proven useful to the study of enzyme functionality under controlled conditions and has led to new technologies. Often, the utility and interest of enzyme-tethered nanostructures lie in how the enzymatic activity is affected by how the enzymes are arranged in space. Therefore, being able to conjugate enzymes to nanostructures while preserving the enzymatic activity is essential. In this paper, we present a method to conjugate single-stranded DNA to the enzyme urease while maintaining enzymatic activity. We show evidence of successful conjugation and quantify the variables that affect the conjugation yield. We also show that the enzymatic activity is unchanged after conjugation compared to the enzyme in its native state. Finally, we demonstrate the tethering of urease to nanostructures made using DNA origami with high site-specificity. Decorating nanostructures with enzymatically-active urease may prove to be useful in studying, or even utilizing, the functionality of urease in disciplines ranging from biotechnology to soft-matter physics. The techniques we present in this paper will enable researchers across these fields to modify enzymes without disrupting their functionality, thus allowing for more insightful studies into their behavior and utility.



\section*{Introduction}


Enzymes are proteins of great interest owing to their ability to catalyze chemical reactions. Without the presence of a catalyst, many biological reactions spontaneously occur with half-times ranging from seconds to millions of years~\cite{snider2000rate}. Enzymes accelerate chemical reactions by orders of magnitude thanks to their substrate-specific active sites~\cite{radzicka1995proficient}, which have evolved to efficiently target unique molecular conformations~\cite{zhang2005enzymes}. Various industries---including pharmaceutical, chemical, and agricultural---use enzymes to decrease production times and costs~\cite{choi2015industrial,li2012technology,liu2013achieve}. Thus, understanding the function of enzymes is necessary for making use of their potential applications in industrial and academic settings. 


Attaching enzymes to synthetic nanostructures has emerged as a promising approach to studying enzyme biophysics and developing new technologies, such as diagnostics. For example, attaching enzymes to magnetic, nanoscale particles has enabled researchers to develop new diagnostic tools for early indication of cancerous cells~\cite{wang2020ph}. Conjugating enzymes to DNA-based nanometer- and micrometer-scale particles has led to the discovery of enhanced diffusion of the conjugate in the presence of the enzyme's substrate, opening new pathways to making enzyme-powered active colloids~\cite{ma2015enzyme,patino2024synthetic}. Finally, DNA origami~\cite{rothemund2006folding, Douglas2009May, Dietz2009Aug, castro2011primer, Gerling2015Mar, baker2018dimensions, sigl2021programmable, hayakawa2022geometrically}, a molecular engineering technology capable of precisely controlling the locations of different enzymes with respect to one another~\cite{funke2016placing, Schnitzbauer2017Jun,kahn2022cascaded, fu2016assembly}, has led to a better understanding of coupled enzymatic reactions~\cite{muller2008dna, sun2017real,kahn2022cascaded} and the roles of motor proteins, like dynein and kinesin, in intercellular transport~\cite{derr2012tug, goodman2014engineering}.

The above applications hinge on the ability to site-specifically functionalize nanostructures or macromolecular complexes with active enzymes. However, general, easy-to-follow protocols for conjugating enzymes to nanostructures that utilize off-the-shelf reagents are limited. Furthermore, the available synthesis approaches often obscure the importance of the various details of the protocol; the choices of the crosslinker-to-enzyme ratio, the solution conditions, and the purification steps used therein are not often investigated or mentioned. Furthermore, since many applications make use of the enzymatic activity of the enzyme, it is critical to retain the enzymatic activity throughout the conjugation procedure, an aspect that is only occasionally discussed~\cite{fu2016assembly}.

\begin{figure} [H]
\includegraphics[]{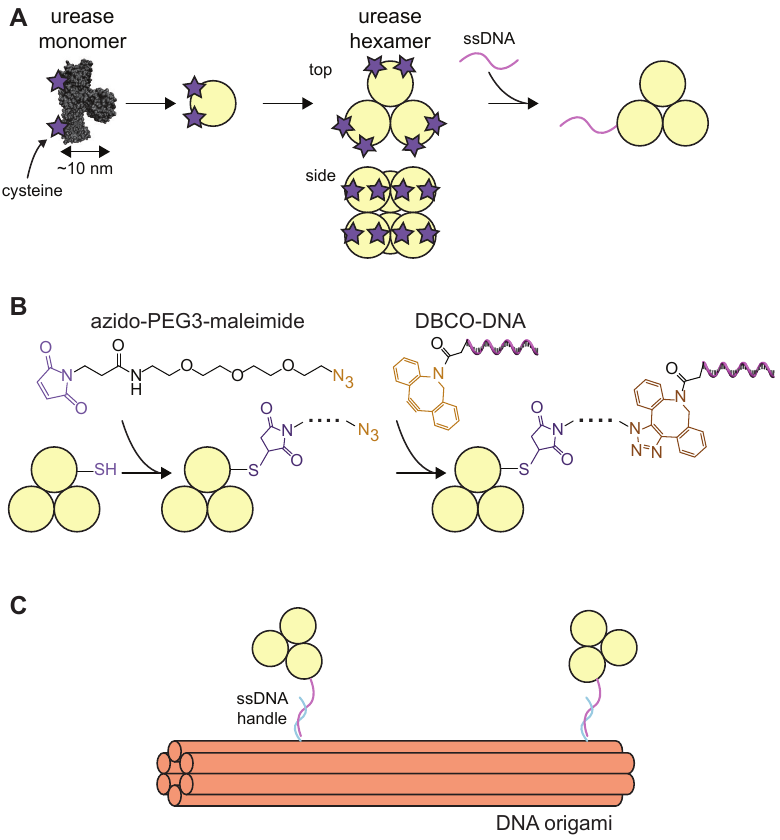}
\caption{{\bf Conjugation scheme.} A) Schematic showing a urease monomer with surface cysteine groups available for binding (purple stars) and six monomers together forming the native hexamer of urease. The top and side views illustrate the hexamer being made up of two trimers stacked on top of one another. Single-stranded DNA can be conjugated to the urease hexamer via the cysteine groups. B) Detailed view of our conjugation scheme. The heterobifunctional crosslinker azido-PEG3-maleimide is incubated with reduced urease, allowing the maleimide (purple) to react with the enzyme’s thiol groups (-SH). A stable thiosuccinimide bond is formed between the maleimide and the thiol. In the second step, DBCO-modified single-stranded DNA is reacted with the urease-azide using ``click'' chemistry. C) Illustration showing DNA-conjugated urease binding to DNA origami via complementary DNA strands (single-stranded DNA handles shown in cyan and magenta).}
\label{fig1}
\end{figure}

In this paper, we present a method to conjugate single-stranded DNA to the enzyme urease while maintaining enzymatic activity, and use the resultant method to bind the DNA-labeled urease to DNA origami with high site-specificity. Our method uses a two-step protocol to attach single-stranded DNA to cysteines on the surface of urease using the thiol-maleimide reaction and click chemistry (Fig~\ref{fig1}A,B). We show proof of a successful DNA-urease conjugation and investigate different factors affecting the conjugation yield. We find that the ratio of crosslinker to enzyme should be comparable to the number of sites on the enzyme complex to maximize yield. Contrary to many other protocols, we also show that the presence of a common reducing agent during the crosslinker incubation can significantly reduce the conjugation yield. Additionally, we find that the presence of salt during the final DNA conjugation step is necessary to achieve appreciable yields, owing to its role in screening the electrostatic interactions between the protein and the DNA. We conclude by demonstrating the ability to bind urease to a six-helix bundle made via DNA origami at user-specified locations with nanometer precision (Fig~\ref{fig1}C), confirmed with transmission electron microscopy. We believe that this experimental system holds promise for elucidating the fundamental behavior and utility of enzymes, and we hope that this paper serves as a useful guide for those hoping to utilize this approach in their field of study. 

\section*{Materials and methods}
\subsection*{Conjugation of single-stranded DNA to urease}
In brief, our synthesis scheme uses a commercially available heterobifunctional crosslinker, azido-PEG3-maleimide (Vector Laboratories, CCT-AZ107), to connect synthetic single-stranded DNA to surface-exposed cysteine (or thiol) groups on the enzyme jack bean urease (Fig~\ref{fig1}A,B). We use urease due to its ubiquity in enzyme studies, its relative durability, and its high turnover rate~\cite{balasubramanian2010crystal}. To conjugate single-stranded DNA to urease, we begin by combining urease with azido-PEG3-maleimide. The maleimide functional group on the crosslinker reacts with the cysteine amino acid side chain on the surface of the urease, forming a thiosuccinimide linkage. We then add single-stranded DNA modified with a dibenzocyclooctyne (DBCO) molecule (DBCO-DNA) to undergo a strain-promoted click reaction with the azide functional group on the other end of the crosslinker~\cite{eeftens2015copper}. Refer to Fig~\ref{fig1}B for a diagram of the conjugation scheme. We verify the conjugation using polyacrylamide gel electrophoresis and quantify the enzymatic activity using colorimetry. 

The first conjugation step is to reduce the thiol groups on the surface of the urease. We dissolve powdered jack bean urease (Canavalia ensiformis (TCI)) to 55~$\upmu$M in 1xPBS buffer at pH~6.7. We add 20~mM of reducing agent tris(2-carboxyethyl)phosphine (TCEP) at pH~7 to the urease at a 10-fold molar ratio of TCEP per urease hexamer, briefly vortex, and allow the solution to incubate in the dark for 1~hour. Meanwhile, we add 500~$\upmu$L of 1xPBS at pH~6.7 to a 0.5~mL Amicon 100-kDa filter and centrifuge the filter at 9,400~x\textit{g} for 8~minutes at room temperature (Fisherbrand accuSpin Micro 17) to equilibrate the column. We use 100-kDa filters because they allow the TCEP (286~Da) to pass through but not the urease hexamers (550~kDa). After the TCEP has incubated with the urease, we add the urease to the prewashed filter and wash it three times with fresh 1xPBS at pH~6.7 each time. Washing the TCEP out of the urease solution is essential for ensuring good conjugation with our crosslinker, as we will show later.

We then react the TCEP-reduced urease with the crosslinker. We start by measuring the urease concentration after washing using a NanoDrop 2000c spectrophotometer (Thermofisher) with the A280 function, dilute it back to 55~$\upmu$M in 1xPBS at pH~6.7, and move it into a 4-mL flat-bottom glass vial with a stir bar. While gently stirring, we slowly add azido-PEG3-maleimide in DMSO at a 5-fold crosslinker to urease hexamer ratio, with final concentrations of 249~$\upmu$M crosslinker and 50~$\upmu$M urease. We cover the tube in foil and allow it to gently stir for 2~hours at room temperature.

Next, we remove unreacted azido-PEG3-maleimide and conjugate the single-stranded DNA to the urease-azide. After two hours, we run the solution through a 0.5-mL, 40-kDa spin desalting column (Thermo Scientific PIA57759) to remove the unbound crosslinker. We collect the washed azide-urease solution and mix it with custom single-stranded DNA (IDT, 5’-TTTTTAACCATTCTCTTCCT-3’, DBCO-modified on the 5' and Cy5-modified on the 3') at a 10-fold DNA to urease hexamer ratio in 1xPBS with a final NaCl concentration of 637~mM (1xPBS has 137~mM NaCl, and we add an additional 500~mM). We incubate this solution overnight in a 40~$^{\circ}$C rotating incubator (Roto-Therm; Benchmark Scientific). After the DNA incubation, we wash the solutions in 0.5-mL Amicon 100-kDa MWCO filters, similar to the TCEP reduction step, to remove unbound DNA (7~kDa), which passes through the filters.

\subsection*{Native polyacrylamide gel electrophoresis}
We verify the success of our conjugation using native polyacrylamide gel electrophoresis (PAGE). Invitrogen NuPAGE 3--8\% Tris-Acetate 1.0~mm mini protein gels were purchased and the manufacturer's instructions were followed to perform PAGE. The running buffer is a 1x solution of Tris-Glycine diluted from a 10x stock (240~mM Tris Base, 1.9~M glycine). The samples are mixed as-is after the conjugation protocol in a 3:1 ratio with 4x loading dye (0.5~M Tris-HCl pH~6.8, 40\% glycerol, 0.1\% (wt/vol) bromophenol blue). We load approximately 5~mg/mL of each sample into the wells. For reference, we load 3~$\upmu$L of NativeMark Unstained Protein Standard (Invitrogen). We run the gel at 150~V for 2.5~hours at 4~$^{\circ}$C. We remove the gel and place it in a Typhoon FLA 9500 laser scanner (GE), then scan it for fluorescence of the Cy5 dye, which is conjugated to DNA. We move the gel into a gel box with Coomassie stain (0.1\% (wt/vol) Coomassie R-250, 30\% methanol, 5\% acetic acid). The gel stains for at least 1~hour, after which time we pour off the Coomassie stain and fill the box with a destaining solution (20\% methanol, 10\% acetic acid). Once the protein bands are clearly visible, we scan the gel in a ChemiDoc MP Imaging System (Bio-Rad).

\subsection*{SDS polyacrylamide gel electrophoresis}
We use SDS-PAGE to complement the results from our native PAGE experiments, ensuring that protein shape does not impact the results. We prepare a 4\% stacking gel on top of an 8\% denaturing, resolving gel between two gel plates. We load approximately 3~mg/mL of each sample into the wells. For reference, we load 3~$\upmu$L of Precision Plus Protein Dual Color Standards (Bio-Rad). We run the gel at 50~V for 30 minutes at room temperature, then increase the voltage to 110~V and run for another hour. When the dye front runs out of the gel, we remove the gel and place it in the above laser scanner, then scan it for fluorescence in the Cy5 range. We move the gel into a gel box with Coomassie stain. The gel stains for at least 1~hours, after which time we pour off the Coomassie stain and fill the box with a destaining solution. Once the protein bands are clearly visible, we scan the gel in a ChemiDoc MP Imaging System.

\subsection*{DNA origami-urease conjugation and negative-stain transmission electron microscopy}
To further verify the conjugation of DNA to urease and to visualize the conjugation of urease to DNA origami, we use transmission electron microscopy (TEM). We mix DNA-conjugated urease at a molar ratio (up to 10-fold) with DNA origami, and dilute the solution to a final DNA origami concentration of 1~nM. We incubate the samples on glow-discharged FCF400-Cu TEM grids (Electron Microscopy Sciences) for 60 seconds at room temperature. We then stain the grids with 2\% uranyl formate solution at 20~mM NaOH for 30 seconds. To image the samples, we use an FEI Morgagni TEM at 80~kV with a Nanosprint5 CMOS camera (AMT).

\subsection*{Enzyme activity assays}
We use a colorimetric assay to measure the enzymatic activity after conjugation. To assess the enzymatic activity of urease, we use phenol red (Sigma-Aldrich 114529) as a means of colorimetric analysis. Urease hydrolyzes urea to produce the weak base ammonia as one of the products. Ammonia increases the pH of the solution, and phenol red changes color from marigold yellow to vivid cerise between pH 6.8 and 8.2. Thus, the activity of urease can be quantified by measuring the change in absorbance of the solution at 560~nm~\cite{okyay2013high}. We prepare samples at a final concentration of 28~$\upmu$M phenol red and 5~mM urea in a 1xPBS buffer at pH~6.7. We add urease to these solutions, either native urease or DNA-modified urease, to a final concentration of 5~nM. We reserve native urease from the conjugation protocol prior to any modifications. For reference, we include blanks with the same phenol red and urease concentrations without urea. Immediately after the addition of urease, we seal the plate with a transparent plate sealer (Thermo Scientific 1424518), and we move the plate to an EPOCH2 microplate reader (BioTek) that takes absorbance measurements of each sample at 560~nm every 60 seconds for up to 90 minutes at 28~$^{\circ}$C. 

\section*{Results and Discussion}
\subsection*{Conjugation of single-stranded DNA to urease}
\subsubsection*{Verification of conjugation using PAGE}
To determine the success of conjugating single-stranded DNA to the surface of urease, we employ native and denaturing PAGE. We run native urease, azide-modified urease, and DNA-modified urease samples from the same conjugation protocol, as well as a protein ladder. The DNA is labeled on its 3’-end with a Cy5 fluorescent dye in order to visualize the DNA in these gels. 

\begin{figure}[H]
\includegraphics[]{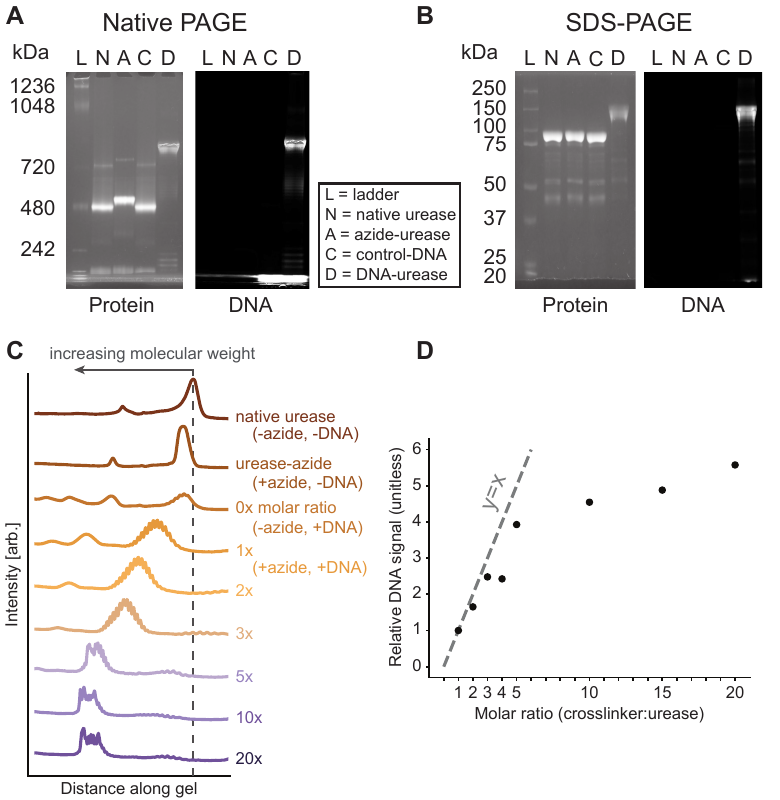}
\caption{{\bf Confirmation of DNA-modification of urease and the role of the molar ratio.}
A) Native PAGE showing the protein stain (left) and Cy5-DNA scan (right) of a reference ladder (L), native urease without DNA (N), azide-modified urease without DNA (A), native urease with DNA (C), and DNA-modified urease (D). There is a clear shift in the primary protein band from the 480~kDa band up to above the 750~kDa band in the ``D'' sample, indicating a larger molecular weight. The DNA signal only appears in the primary protein band of the ``D'' sample, indicating the presence of DNA. B) Denaturing PAGE showing the same samples as in A), with an appropriate protein ladder. We make the same observations as in A), with an increase in the molecular weight of the primary protein band of ``D'' and the presence of DNA signal in the same band. C) Plot of the average intensity along the lanes of a protein-stained native PAGE gel for urease samples of various molar ratios of crosslinker to urease hexamer. As the molar ratio increases, the molecular weight of the sample increases until after 10x when it saturates. The dotted line corresponds to the position of urease in its native, hexameric form. D) Plot showing the relative DNA signal in the conjugation sample as a function of the molar ratio, normalized to the 1x condition. We define relative DNA signal as the blank-subtracted signal intensity of the main band in the DNA scan divided by the blank-subtracted signal intensity of the main band in the protein scan, normalized to the 1x condition ($\frac{\sfrac{I_{\textrm{DNA}} - I_0}{I_{\textrm{protein}} - I_0}}{\sfrac{I^{1x}_{\textrm{DNA}} - I_0}{I^{1x}_{\textrm{protein}} - I_0}}$). The dashed line shows y=x, highlighting that the ratio increases linearly with a slope of around 1 for molar ratios up to 5x.}
\label{fig2}
\end{figure}

We infer the success of our conjugation, as well as the number of conjugated DNA molecules, from the band position and intensity. Figure~\ref{fig2}A shows a scan of the native PAGE gel of our samples for one experimental condition. The primary protein bands of the native-urease-without-DNA, azide-modified-without-DNA, and native-urease-with-DNA samples (Fig~\ref{fig2}A ``N'', ``A'', ``C'', respectively) run similarly to the 480-kDa band in the ladder. This molecular weight is consistent with the previous observations of native urease with molecular weights between 480~kDa and 590~kDa that are indicative of a native hexameric form~\cite{sumner1938molecular, dixon1980jack}. There is a slight shift up in the azide-modified urease band compared to the native urease band, which may be due to the non-negligible molecular weight of the crosslinker (369.37~Da). The primary protein band of the DNA-modified urease runs above the 720~kDa band in the ladder (Fig~\ref{fig2}A ``D''). We attribute this apparent increase in the molecular weight of the protein to the conjugated DNA, which itself has a molecular weight of about 7~kDa. It is also worth noting that the shape and charge of molecules affect how they run through the gel, making it difficult to predict how the DNA-enzyme conjugate will run compared to globular proteins, for example.

To confirm the conjugation, we also scan the same gel for Cy5 fluorescence, shown as the DNA scan in Fig~\ref{fig2}A. The DNA-modified urease band shows a strong fluorescent signal, while the other urease bands show little to no signal. We thus conclude that we have successfully conjugated single-stranded DNA to urease.

Given that the tertiary and quaternary structures of proteins can change and influence native PAGE results, we also run these samples in a denaturing gel that breaks urease oligomers down to their monomeric form and linearizes them. The only reason the protein band would shift between samples in a denaturing gel is due to the attachment of another molecule since the amino-acid sequence is the same for all samples. Figure~\ref{fig2}B shows the protein and DNA scans for an SDS-PAGE experiment. Again, there is a shift in the bands of the DNA-modified sample compared to the bands in the native and azide-modified samples. There is also a clear fluorescent signal in the largest, most prominent shifted band in the Cy5 channel, indicating that DNA is conjugated to the proteins in that sample. By eliminating protein shape as a variable in how the gels run, these results confirm the covalent conjugation of DNA to urease.

In addition to confirming the conjugation of DNA to urease, we use PAGE gels to quantify the influence of some of the most important variables affecting the final yield. In particular, we test the effects of the molar ratio of crosslinker to urease hexamer in the first step, the amount of salt in the final reaction step, and the presence of a reducing agent in the crosslinker incubation. The amount of DNA bound to the urease is inferred by comparing how far the protein bands of different samples run in the gel, as well as the ratio of the intensity of the conjugate bands from the protein and DNA dyes.

\subsubsection*{Effect of crosslinker molar ratio on conjugation yield}
The molar ratio of crosslinker to urease hexamer has a large effect on the success of conjugation, with too little crosslinker negatively impacting the final yield. We consider the molar ratio of the crosslinker compared to the urease hexamer. Figure~\ref{fig2}C shows intensity scans of a native PAGE gel for urease samples incubated with various molar ratios of crosslinker during the conjugation protocol. As expected, native urease, azide-modified urease, and urease mixed with pure DMSO in place of crosslinker (0x) all have comparatively similar molecular weights, since no DNA is introduced or able to bind to these enzymes. As the molar ratio increases from 1x to 10x, there is an increase in the molecular weight of the primary protein band. We attribute this observation to an increasing amount of crosslinker, and thus DNA, binding to the urease. Above a molar ratio of 10x, the band position appears to saturate.

We also consider the signal intensities within the PAGE bands and what they can tell us about the conjugation yield. Figure~\ref{fig2}D shows a plot of the ratios of the integrated densities of the DNA signal to the protein signal within the main band of each molar-ratio condition, normalized to the ratio at 1x. We see that these relative signals follow the same trend as the running distances, with an increasing shift in the band position corresponding to an increasing relative signal of DNA to protein. Moreover, we observe that the relative DNA to protein signal increases roughly linearly with increasing molar ratio up to 5x before saturating around 10x.

Our results indicating an optimal crosslinker to urease hexamer molar ratio of about 10x are consistent with a rough estimate of the number of accessible thiol groups. Based on its structure, we estimate that a urease hexamer has roughly 12 potential thiol groups on its surface for labeling \cite{balasubramanian2010crystal}. This number is comparable to the molar ratios at which we observe a plateau in the shifts of the DNA-labeled protein bands. Therefore, given the linear dependence of the relative DNA signal and the optimal molar ratio of 10x, we suspect that the yield of the thiol-maleimide reaction in the first step of our conjugation scheme is quite high.

\subsubsection*{Increasing salt concentration improves the conjugation yield}
\begin{figure}[H]
\includegraphics[]{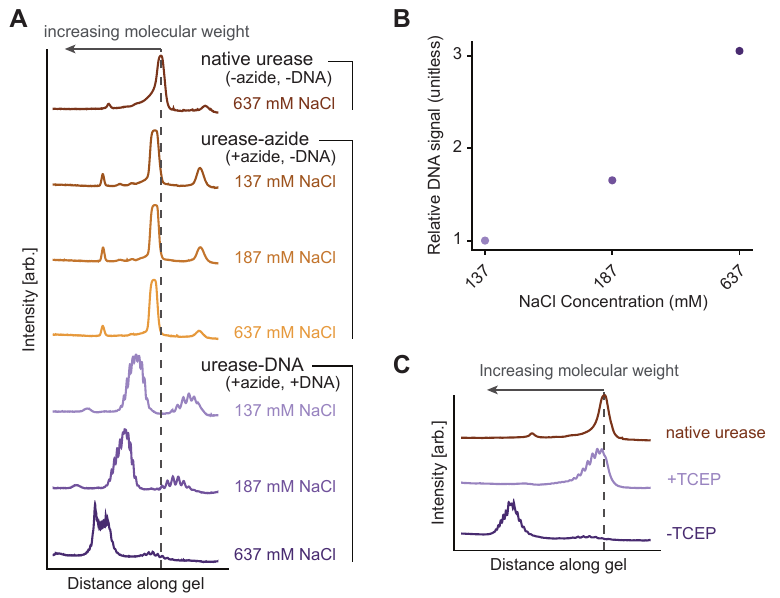}
\caption{{\bf PAGE gel analysis of urease conjugation for various salt conditions and removal of reducing agent.} A) Plot of the average intensity along lanes of a protein-stained native PAGE gel for conjugation samples with varying amounts of NaCl added during the DNA reaction step. At increasing NaCl concentrations, the positions of the urease-azide bands remain similar while the urease-DNA bands show increasing molecular weight that we attribute to higher degrees of DNA labeling. B) A plot showing the relative DNA signal increase in the labeling yield as a function of the NaCl concentration. We define relative DNA signal similarly to in Fig~\ref{fig2}. C) Same plot in A) but for conjugation samples with TCEP not washed out (+TCEP) and with TCEP washed out (-TCEP) immediately following the TCEP incubation step in the protocol. The main protein band of the -TCEP sample is of a higher molecular weight than the native and +TCEP samples, indicating successful conjugation of DNA to urease.}
\label{fig3}
\end{figure}

We demonstrate that increasing the salt concentration in the final DNA incubation reaction increases the conjugation efficiency of the second step of our scheme. Figure~\ref{fig3}A shows the protein bands for various urease samples at 137~mM, 187~mM, and 637~mM NaCl in the final solution. All conjugations referenced henceforth use a 5x and 10x molar ratio of crosslinker and DBCO-DNA  to urease hexamer, respectively. When no DBCO-DNA is added to the final reactions, there is no difference in molecular weight between the samples with different NaCl concentrations. However, when DBCO-modified DNA is added, there is a clear increase in the molecular weight of the samples as the NaCl concentration increases. The plot in Fig~\ref{fig3}B shows the ratios of the DNA signal to the protein signal within the main band of each salt condition, normalized to the ratio at 137~mM NaCl. The conjugation yield increases by a factor of roughly 1.6, going from 137~mM NaCl to 187~mM NaCl, and by approximately three-fold from 137~mM NaCl to 637~mM NaCl. Therefore, we conclude that the NaCl concentration plays an important role in determining the success and yield of the DNA-to-urease conjugation step.

We attribute the increase in conjugation efficiency to the salt-dependent screening of the negative charges on the DNA and the enzyme. Both DNA and urease are negatively charged at neutral pH, given that urease’s isoelectric point is about 5.1~\cite{sumner1929isoelectric}. We hypothesize that the addition of salt screens the electrostatic repulsion between the two negatively charged molecules, enabling them to come closer together and thus resulting in more conjugation events. Other protocols conjugating DNA to urease do not include additional salt and are mostly done in 1xPBS, which has a NaCl concentration of 137~mM and a KCl concentration of 2.7~mM~\cite{chang2017detection, wang2020ph, patino2024synthetic}. Therefore, compared to existing protocols in the literature, our findings suggest that the conjugation efficiency can be improved roughly three-fold by using monovalent salt concentrations above 600~mM.

\subsubsection*{The presence of TCEP significantly reduces crosslinking efficiency}
Finally, we report a subtle but important detail in the protocol to achieve higher yield: The presence of the reducing agent tris(2-carboxyethyl)phosphine (TCEP) significantly impedes the conjugation of DNA to urease. We incubate urease with TCEP prior to incubation with the crosslinker because it cleaves any disulfide bonds at the urease surface, thus freeing up the thiol groups for conjugation with the maleimide-end of the crosslinker. After this incubation, washing TCEP out of the solution with 1xPBS using a molecular weight cutoff filter dramatically improves the conjugation success. Using PAGE gel analysis, we observe that the urease from the reaction with TCEP removed (-TCEP) is of a higher molecular weight than native urease and urease from the reaction with TCEP not removed (+TCEP) (see Fig~\ref{fig3}C). These conjugations use a 5x molar ratio of crosslinker. Indeed, the yield increases roughly four-fold when we remove TCEP compared to when we do not remove TCEP, as determined by the relative DNA signals.

We attribute the importance of removing TCEP to the fact that the presence of TCEP may reduce the ability of the maleimide-functionalized crosslinker to react with the thiol groups on the urease. This result is consistent with a previous report demonstrating that TCEP can react with maleimide groups to form non-reactive byproducts~\cite{kantner2016characterization}. It is also consistent with a report detailing how the presence of TCEP significantly reduced the conjugation efficiency of maleimide labeling~\cite{getz1999comparison}. However, many publications do not mention that this consideration should be made~\cite{scales2006fluorescent, cumnock2013trisulfide, liu2014intracellular}, and some online protocols~\cite{fisherprotocol} state that it is unimportant to remove TCEP after reducing the disulfide bonds, which is inconsistent with our observations. Our experiments clearly show that removing TCEP is a necessary step in achieving successful urease-DNA conjugation using a maleimide functional group.

\subsection*{Enzyme-activity assay}
We verify that the enzymes remain active after DNA conjugation using phenol red. The absorbance of phenol red at 560~nm increases as the pH of the solution increases from 6.8 to 8.2. When urease catalyzes the decomposition of urea, it increases the pH of its solution by creating ammonia. Therefore, an increase in the absorbance at 560~nm by solutions containing urease, urea, and phenol red indicates enzymatic activity~\cite{okyay2013high}.

We plot the absorbance change of native urease and DNA-modified ureases incubated with varying molar ratios of crosslinker. To quantify the enzymatic activity from these data, we fit a line to the raw absorbance data between 3--10 minutes. We take the slope of each line to be the reaction rate when substrate is saturating. To these values, we fit the Hill equation (Equation~\ref{eq:1}), a general version of the Michaelis-Menten equation~\cite{goutelle2008hill}:
\begin{equation} \label{eq:1}
    v = \frac{V_\textrm{max} [S]^{n}}{K_{\rm M}^n + [S]^n},
\end{equation}
where $v$ is the reaction rate, $V_\textrm{max}$ is the maximal rate of reaction, $[S]$ is the substrate concentration, $n$ is the Hill coefficient, and $K_{\rm M}$ is the Michaelis constant. By fitting this equation to our data, we extract best-fit values for $n$, $K_{\rm M}$, and $V_\textrm{max}$. We stress that $v$ and therefore $V_\textrm{max}$ in our analysis do not have direct biochemical interpretations because they result from measurements of the rate of change of the absorbance of phenol red and not the rate of change of a molar concentration of reactants or products. 

\begin{figure}[H]
\includegraphics[]{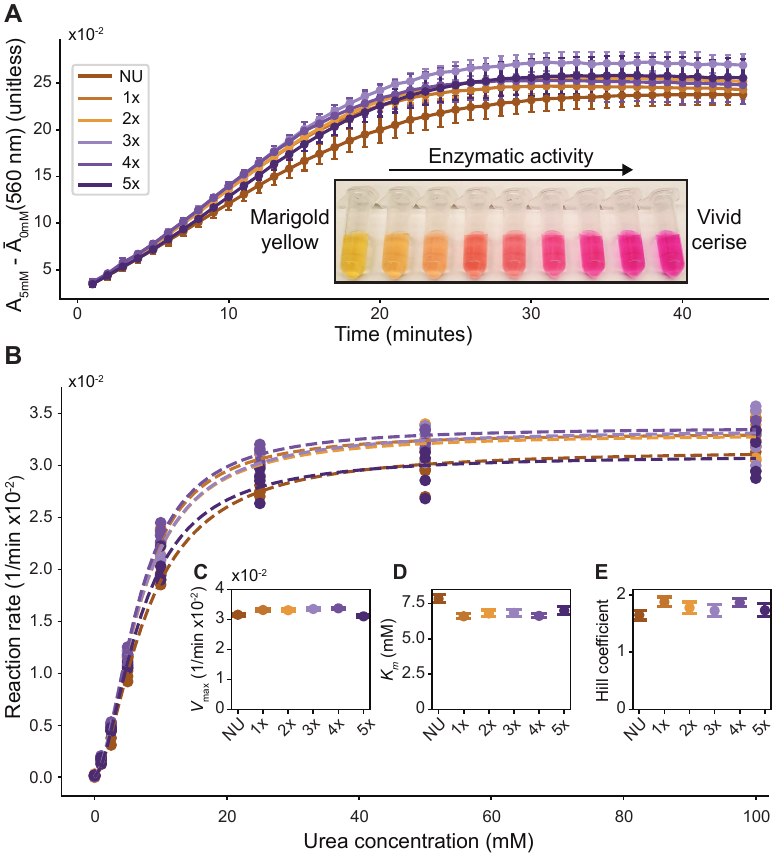}
\caption{{\bf Confirmation of enzymatic activity using a colorimetric assay.}
A) Absorbance at 560~nm of native urease (NU) and DNA-modified ureases incubated with varying molar ratios of crosslinker to urease hexamer during the conjugation, in the presence of 28~$\upmu$M phenol red and 5~mM urea. These data are averaged across four repeated experiments. Error bars show the standard deviations calculated from these four measurements. From each dataset, we subtract the mean absorbance from controls of the same urease and phenol red concentrations with no urea present. The inset shows a photograph of the absorbance change of phenol red as the pH increases from around 6.8 (marigold yellow) to around 8.2 (vivid cerise). B) Reaction rates of urease samples in the presence of various urea concentrations. We calculate the reaction rate as the slope of a line fit to the raw absorbance between 3--10~min. Each condition at each urea concentration has four points from four repeated experiments. Dotted lines represent fits of the Hill equation (Equation~\ref{eq:1}) to the four points per urea concentration. C) The maximum rate of reaction $V_\textrm{max}$ for each condition, taken from the fit in B). Error bars represent standard errors taken from the fit. D) The Michaelis constant $K_{\rm M}$ for each condition. E) The Hill coefficient $n$ for each condition.}
\label{fig4}
\end{figure}

We find that the conjugation of DNA to urease does not change the enzymatic activity and that the activity is largely independent of the DNA labeling efficiency. Figure~\ref{fig4}A shows an example of the absorbance over time for native urease and DNA-modified ureases incubated with varying molar ratios of crosslinker in 5 mM urea. The slopes of the linear portions of these data are the reaction rates, which we plot as a function of the urea concentration in Figure~\ref{fig4}B. We fit Equation~\ref{eq:1} to these data and extract the parameters plotted in Figure~\ref{fig4}C-E. There is no systematic trend in these parameters, indicating no change in the enzyme performance as it relates to crosslinker molar ratio. Our measured values of $n$ are close to previously reported values for jack bean urease of 2.19~\cite{milo2021small}, while our values of $K_{\rm M}$ are within the range of reported values, 3.0--12~mM, for various phosphate buffer concentrations~\cite{krajewska1999effect}. It should be noted that these measured parameters are highly sensitive to enzyme purity, buffer type, pH, and temperature, and are thus difficult to directly compare between experiments not performed in the same aqueous environment~\cite{krajewska2009ureases}. 

Our results contrast with other published experiments that demonstrate a decrease in activity for other enzymes as the number of labeled crosslinkers per protein increases~\cite{fu2016assembly}. We speculate that we may not see this trend using our protocol and the enzyme urease for a few reasons. First, the prior experiments~\cite{fu2016assembly} did not use urease (instead using glucose oxidase, horseradish peroxidase, lactate dehydrogenase, alkaline phosphatase, and glucose-6-phosphate dehydrogenase), and we hypothesize that the impact of labeling on the activity may depend sensitively on the details of the active site and where labeling occurs relative to it. Second, we hypothesize that the specific chemical group used for conjugation may also influence the resultant activity. For example, our method targets thiol groups on the protein surface, while the previous study targets amine groups using the crosslinker succinimidyl 3-(2-pyridyldithio)propionate (SPDP). For these reasons, we argue that it is important to directly confirm the activity of any new combination of enzyme and crosslinker, and that no single conjugation protocol is likely to be a magic bullet.

\subsection*{Conjugating urease to DNA origami}

Because the ultimate motivation for developing this synthesis scheme is to utilize DNA hybridization to make enzyme-tethered DNA nanostructures, we finally confirm that our DNA-enzyme constructs can be site-selectively bound to DNA origami. 

\begin{figure}[H]
\includegraphics[]{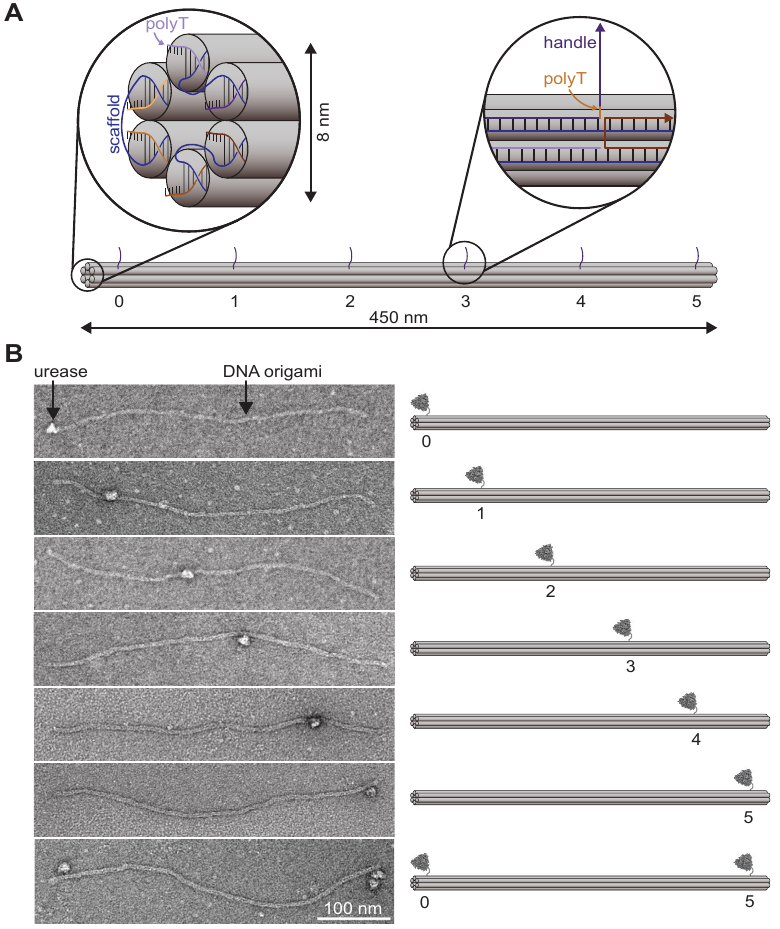}
\caption{{\bf Site-specific binding of urease to DNA origami rod in TEM.}
A) Illustration showing the design of the six-helix bundle DNA origami. The rod-like structure has a length of approximately 450~nm and a diameter of approximately 8~nm. We show the organization of DNA, including the scaffold (blue), within the six-helix bundle at the cap of the structure as well as an example routing schematic of one of the handle DNA strands (purple) extruding from the structure. B) TEM micrographs showing urease bound to DNA origami rods at various locations along their lengths. Single-stranded DNA was pulled out of the DNA origami rod at regular intervals to allow DNA-modified urease to bind via hybridization. The DNA-modified urease is mixed with each DNA origami sample at a 10x molar ratio. The image with the scale bar shows a DNA origami rod with two handles for the enzyme to conjugate to at each end.}
\label{fig5}
\end{figure}

We design and fold the DNA origami into a six-helix bundle~\cite{mathieu2005six}. The resulting rod-like structure is folded from an 8064-nucleotide circular DNA scaffold and has a diameter of approximately 8~nm and a length of around 450~nm. Figure~\ref{fig5}A shows a diagram of the six-helix bundle, highlighting the DNA organization within the structure. Five thymines extend from the ends of the helices and prevent base-pair stacking between six-helix bundles. For conjugating urease to this DNA origami, we extrude single-stranded DNA from the structure at specific locations. We design the strands to come out at roughly 85-nm intervals at six locations along the full length of the six-helix bundle. The strands, with a binding domain of sequence 5'-AGGAAGAGAATGGTT-3', are complementary to the strand conjugated to the urease. Five thymines precede the binding domain toward the 5'--end to act as a spacer between the handle and the DNA origami structure. By mixing urease with the complementary DNA strand attached in a 10:1 ratio of urease to DNA origami, the two join together via DNA hybridization. See Supporting Information File S1 for detailed protocols for DNA-origami design, folding, purification, and imaging.

Using TEM, we see direct visual evidence of enzymes bound to DNA origami. Figure~\ref{fig5}B shows TEM micrographs of DNA-origami rods with DNA-modified urease bound to them at regular intervals along the length of the rod. The bottom image in the series shows urease bound at each end of the DNA origami. Of the 305 DNA origami rods we imaged, we found urease bound to 180 of them, a binding efficiency of approximately 60\%. For all DNA origami with a bound urease, no more than one urease was bound to a single structure and we did not detect any off-target bindings. These observations are consistent with the fact that each DNA origami was folded to possess at most one of six possible anchor strands, except for the last scenario in Fig~\ref{fig5}B which allows for two bindings.
 
We speculate that the binding efficiency may be hampered due to the inactivation of binding sites or slow kinetics. While we wash the unreacted DNA oligomers out of the solution at the end of the conjugation (as outlined in the first section of the Methods), it is possible that some free DNA strands remain and can compete to bind to the DNA origami. It is also possible that the binding of urease-DNA to DNA origami is slow and that we do not wait long enough for all of the potential binding sites to be occupied at the concentrations we use for incubation. The samples are mixed with a final DNA-origami concentration close to 1~nM and a final urease concentration close to 10~nM. Calculation of the dissociation constant for our DNA binding domain at room temperature, 5~mM MgCl\textsubscript{2}, and 50~mM NaCl yields approximately 1~fM~\cite{zadeh2011nupack}. At concentrations less than 1~fM we would expect to see complete dissociation of these strands, implying that our mixing concentrations are high enough. We allow our combined samples to incubate for 45 minutes prior to negative-staining, therefore if the binding is slow then increasing the incubation time may improve the binding efficiency. To test how temperature affects tethering efficiency, we ramped a solution of DNA origami and DNA-conjugated urease from 37~$^{\circ}$C to 25~$^{\circ}$C over the course of one hour. We found that, of the 220 rods imaged, urease was bound to 133 of them, a binding efficiency of about 60\%. This temperature ramp over this amount of time did not improve binding efficiency from a similar time-span at room temperature.

\section*{Conclusion}

In summary, we developed a protocol to conjugate urease to DNA-origami nanostructures with high site-specificity while maintaining enzymatic activity. We showed that the crosslinker azide-PEG3-maleimide offers a way to modify urease with an azide which can be reacted with DBCO-modified DNA to form DNA-labeled urease. We found that a ten-fold molar ratio of crosslinker to urease hexamer results in the highest labeling yield. We also found that the presence of a popular reducing agent can impede conjugation of the crosslinker, contrary to some conjugation protocols~\cite{scales2006fluorescent, cumnock2013trisulfide, liu2014intracellular, fisherprotocol}. We showed that increasing the monovalent salt concentration during the DNA reaction step can increase conjugation yield. Critically, our conjugation protocol does not reduce the enzymatic activity of the urease compared to its native state. We are thus able to bind enzymatically-active urease to the surface of DNA-origami nanostructures with great site-specificity. 

We conclude by highlighting that our method is one among many published approaches to labeling proteins with DNA~\cite{cremers2019efficient, trads2017site}. These methods follow a variety of strategies, including crosslinking~\cite{wang2020ph, cremers2019efficient, patino2024synthetic}, unnatural amino acid modification~\cite{lee2019site, lang2014cellular, nikic2015genetic}, N-terminal modification~\cite{gilmore2006n, dixon1984n}, and others. The technique we employ uses a two-step protocol to attach single-stranded DNA to cysteines on the surface of urease using the thiol-maleimide reaction and click chemistry (Fig~\ref{fig1}A,B). We choose this approach because it makes use of easily-accessible, off-the-shelf reagents and can achieve high reaction yields. Because our approach does not require any specialized expertise in chemical synthesis and uses only commercially available reagents, we hope it will be accessible to a broad audience, including scientists from outside the chemical and life sciences. 
Furthermore, we are hopeful that our thorough examination of the various factors that influence the success and yield of the conjugation steps will enable researchers to utilize this experimental system, and hope that this protocol offers guidance in this regard.

\section*{Supporting Information}
\noindent \textbf{File S1. Detailed methods.} Supporting Information text provides detailed protocols for DNA-origami design, folding, purification, and imaging.  

\section*{Acknowledgments}
We thank Kayla Cerri of the Krauss Lab for invaluable assistance with MALDI-TOF analysis of our DNA-enzyme conjugates. We thank Zachary Curtis from the Bisson Lab for assistance with operating the plate reader. We thank Bryan Gworek for assistance in debugging the early stages of the project. TEM samples were prepared and imaged at the Brandeis Electron Microscopy Facility. 



%
%
%

\end{document}


\section*{Supporting information}

\subsection*{DNA origami techniques}
\paragraph*{Designing DNA origami}\label{subsec:design}
The full design for the six-helix bundle is available on Nanobase as structure 249~\cite{poppleton2022nanobase}.

\paragraph*{Folding DNA origami}\label{subsec:folding}
Each DNA origami particle is folded by mixing 50 nM of p8064 scaffold DNA, which has 8064 nucleotides (Tilibit), and 200 nM each of staple strands with folding buffer and annealed through a temperature ramp starting at 65~$^{\circ}$C for 15 minutes, then 54 to 51 ~$^{\circ}$C, $-1~^{\circ}$C per hour. Our folding buffer contains 5 mM Tris Base, 1 mM EDTA, 5 mM NaCl, and 5 mM MgCl\textsubscript{2}. We use a Tetrad (Bio-Rad) thermocycler for annealing the solutions. 

\paragraph*{Agarose gel electrophoresis}\label{subsec:electrophoresis}
To assess the outcome of folding, we perform agarose gel electrophoresis. Gel electrophoresis requires the preparation of the gel and the buffer. The gel is prepared by heating a solution of 1.5\% wt/wt agarose, 0.5x TBE to boiling in a microwave. The solution is cooled to 60~$^{\circ}$C. At this point, we add MgCl\textsubscript{2} solution and SYBR-safe (Invitrogen) to adjust the concentration of the gel to 5.5~mM MgCl\textsubscript{2} and 0.5x SYBR-safe. The solution is then quickly cast into an Owl B2 gel cast, and further cooled to room temperature. The buffer solution contains 0.5x TBE and 5.5~mM MgCl\textsubscript{2}. Agarose gel electrophoresis is performed at 90~V for 1.5 hours at 4~$^{\circ}$C. The gel is then scanned with a Typhoon FLA 9500 laser scanner.

\paragraph*{Gel purification and resuspension}\label{subsec:purification}
After folding, DNA-origami particles are purified to remove all excess staples and misfolded aggregates using gel purification. The DNA origami are run through an agarose gel (now at a 1xSYBR-safe concentration for visualization) prepared using a custom gel comb, which can hold around 4~mL of solution per gel. We use a black light box (Hall Productions BL1012) to identify the gel band containing the folded DNA origami. The folded origami band is then extracted using a razor blade and cut into pieces. We place the gel pieces into a Freeze 'N Squeeze spin column (Bio-Rad), freeze it in a --80~$^\circ$C freezer for 30 minutes, thaw at room temperature, and then spin the solution down for 5 minutes at 13,000~x\textit{g}. 

Next, we concentrate the solution through ultrafiltration \cite{wagenbauer_how_2017}. First, a 0.5-mL Amicon 100 kDA ultrafiltration spin column is equilibrated by centrifuging down 0.5~mL of the folding buffer at 5,000~x\textit{g} for 7 minutes. Then, the DNA-origami solution is added up to 0.5~mL and centrifuged at 14,000~x\textit{g} for 15 minutes. Finally, we flip the filter upside down into a new Amicon tube and spin down the solution at 1,000~x\textit{g} for 2 minutes. The concentration of the DNA origami is measured using a Nanodrop (Thermofisher), assuming that the solution consists only of well-folded particles that are each 8064 base pairs.

\paragraph*{Negative stain TEM}\label{subsec:TEM}
We first prepare a solution of uranyl formate (UFo). Millipore water is boiled to deoxygenate it and then mixed with uranyl formate powder to create a 2\% wt/wt UFo solution. The solution is covered with aluminum foil to avoid light exposure, then vortexed vigorously for 20 minutes. The solution is filtered using a 0.2~$\upmu$m filter. The solution is divided into 0.2-mL aliquots, which are stored in a --80~$^\circ$C freezer until further use.

Prior to each negative-stain TEM experiment, a 0.2-mL aliquot of UFo is taken out from the freezer to thaw at room temperature. We add 4~$\upmu$L of 1 M NaOH to precipitate the UFo and vortex the solution vigorously for 15 seconds. The solution is centrifuged at 4~$^\circ$C and 16,000~x\textit{g} for 8 minutes. We extract 170~$\upmu$L of the supernatant for staining and discard the rest. 

The EM samples are prepared using FCF400-Cu grids. We glow discharge the grid prior to use at --20 mA for 30 seconds at 0.1~mbar, using a Quorum Emitech K100X glow discharger. We place 4~$\upmu$L of the sample on the grid for 1 minute to allow adsorption of the sample to the grid. During this time 5~$\upmu$L and 18~$\upmu$L droplets of UFo solution are placed on a piece of parafilm. After the adsorption period, the remaining sample solution is blotted on a Whatman filter paper. We then touch the carbon side of the grid to the 5~$\upmu$L drop and blot it away immediately to wash away any buffer solution from the grid. This step is followed by picking up the 18~$\upmu$L UFo drop onto the carbon side of the grid and letting it rest for 30 seconds to deposit the stain. The UFo solution is then blotted to remove excess fluid. Grids are dried for a minimum of 15 minutes before insertion into the TEM.

We image the grids using an FEI Morgagni TEM operated at 80 kV with a Nanosprint5 CMOS camera. The microscope is operated at 80 kV and images are acquired between x8,000 to x28,000.